\documentclass[aps,prd,twocolumn,groupedaddress,showpacs,preprintnumbers,amsmath,amssymb,floatfix]{revtex4}

\usepackage{graphicx}
\usepackage{subfig}
\usepackage{dcolumn}
\usepackage{float}
\usepackage{bm}
\usepackage{enumerate}

\usepackage[english]{babel}
\usepackage{amsfonts}
\usepackage{amsthm}
\usepackage{amssymb}
\usepackage{calc}
\usepackage{ifthen}

\newcommand{\GL}{Ginzburg-Landau }

\theoremstyle{definition}
\ifthenelse{\not\isundefined{\thechapter}}{

}{}

\newtheorem*{huom*}{Huomautus}
\newtheorem*{seuraus*}{Seuraus}
\newtheorem*{esim*}{Esimerkki}

\newcommand{\set}[1]{\mathrm{#1}}
\newcommand{\spac}[1]{\set{#1}}

\newcommand{\density}[1]{\mathcal{#1}}

\renewcommand{\vec}[1]{\boldsymbol{#1}}
\newcommand{\half}{\tfrac{1}{2}} 
\newcommand{\abs}[1]{\lvert{#1}\rvert} 

\newcommand{\norm}[1]{\lVert#1\rVert}

\newcommand{\adjoint}[1]{#1^\dagger}

\newcommand{\diff}{\mathrm{d}}

\newcommand{\scalarprod}{\cdot}
\newcommand{\crossprod}{\times}

\DeclareMathOperator{\defin}{\mathrm{:=}}

\begin{document}

\title{Stability of topological solitons in modified two-component Ginzburg-Landau model}

\author{Juha J\"aykk\"a}
\email{juolja@utu.fi}
\affiliation{
Department of Physics and Astronomy,
University of Turku,
FI-20014 Turku, Finland
}

\date{\today}

\begin{abstract}
  We study the stability of Hopfions embedded in a certain modification \GL model of two equally charged
  condensates. It has been shown by Ward [Phys.\ Rev.\ D{\bf 66}, 041701(R) (2002)] that certain modification
  of the ordinary model results in system which supports stable topological solitons (Hopfions) for some
  values of the parameters of the model. We expand the search for stability into previously uninvestigated
  region of the parameter space, charting an approximate shape for the stable/unstable boundary and find that,
  within the accuracy of the numerical methods used, the energy of the stable knot at the boundary is
  independent of the parameters.
\end{abstract}

\maketitle

\section{\label{sec:intro}Introduction}
Topological solitons, be it vortices, knots, instantons or other objects, enjoy widespread interest within
many fields of physics, perhaps most notably in the fields of particle physics and condensed matter, where
these objects invariably occur as solutions of the field equations. In condensed matter physics, topologically
stable vortices are also a routinely seen in experiments. Therefore, it is crucial to understand the basic
properties of topological solitons in the \GL and related models. It is against this background that we have
studied the static \GL model, also known as the Abelian Higgs model. Some years ago, it was demonstrated that the
archetypal 3D classical field model supporting topologically stable {\it closed} vortices, the Faddeev-Skyrme
model \cite{Faddeev:1975tz, Faddeev:1979aa, Faddeev:1997zj, Battye:1998pe, Battye:1998zn, Hietarinta:1998kt,
  Hietarinta:2000ci, Hietarinta:2003vn, Adam:2006bw}, can be embedded in the \GL model by a change of
variables \cite{Babaev:2001zy}. There was also an earlier work, where the FS model was obtained from \GL model
in a derivative expansion \cite{Hindmarsh:1992yy}, but this method does not allow for investigation of
solitons in \GL model since it does not correspond to any parameter limit. It was further conjectured in
Ref.~\cite{Babaev:2001zy} that the two-component \GL model should, due to this embedding, also support the
same topological structures as the FS model does, but more recent studies do not support this conjecture
\cite{Ward:2002vq, Jaykka:2006gf}. However, Ward found \cite{Ward:2002vq} that by modifying the model
suitably, stable closed vortices appear as minimum energy configurations of the theory. We expand on Ward's
work and find the stable/unstable boundary in the $(\kappa, \eta)$ parameter space of the model. We also find
that the energy of the stable minimum energy configuration along the said boundary is constant (to within the
accuracy of the methods used). The results presented here should also be useful in constructing such initial
configurations in \GL model which relax into a local, non-zero energy minimum instead of the global one and
thus provide a way to construct knotted solitons in the \GL model.

\section{\label{sec:model}The model}
The static Abelian Higgs model with two charged Higgs bosons is mathematically the same as the \GL model
with two flavors of Cooper pairs or super-fluids. The paper will use the following notations.  The indices run
as follows: $j,k,l \in \{1,2,3\}$, $\alpha \in \{1,2\}$, $\mu,\nu,\iota \in \{0,1,2,3\}$ and the fields are
$\Psi = (\psi_1 \quad \psi_2)^T$, $F_{\mu\nu} \equiv \partial_\mu A_\nu - \partial_\nu A_\mu$, $\vec{B} =
\epsilon^{jkl}\partial_k A_l$ and the gauge-covariant derivative has the form $D_\mu \equiv \partial_\mu - i
\tfrac{2e}{\hbar c}A_\mu$; when working in three dimensions ($\mu \in \{1,2,3\}$) we will also write $\vec{D}
\equiv \nabla - i \tfrac{2e}{\hbar c} \vec{A}$.  With these notations, the standard Lagrangian density of the
two-component \GL model can be written as
\begin{align}
  \label{eq:LtcGLSI}
     {\density L} &=
     \tfrac{\hbar^2}{2m_\alpha} \norm{D_\mu \psi_\alpha}^2 
     +V\bigl(\psi_{1},\psi_{2}\bigr) - \tfrac{1}{4}F_{\mu\nu}F^{\mu\nu},
  \intertext{which gives the static energy density}
  \label{eq:EtcGLSI}
   {\density E} &=
   \tfrac{\hbar^2}{2m_\alpha} \norm{\vec{D} \psi_\alpha}^2 
    +V\bigl(\psi_{1},\psi_{2}\bigr) + \tfrac{1}{2\mu_0}\norm{\vec{B}}^2,
\end{align}
where we have used SI units. The form of the potential is not very important as long as it maintains the
$SU(2)$ symmetry of $\Psi$ and enforces the condition $\norm{\Psi}=$~constant~$\ne 0$ at some limit of the
parameters of the potential; here we have used
\begin{align}
  \label{eq:potential}
  V(\psi_{1},\psi_{2}\bigr) &= \half \eta (\abs{\Psi}^2-1)^2.
\end{align}
The electric coupling constant displays an explicit factor of 2 due to the
interpretation of the \GL model as a superconductor, where the $\psi_\alpha$ become Cooper
pairs. For computational purposes, it is practical to use natural units, where $\hbar = c = \mu_0 = 1$ and
rescale the fields by $\psi_\alpha \rightarrow \psi_\alpha \sqrt{m_\alpha}$. For flexibility, we retain a
freely selectable electric charge but replace $2e \rightarrow g$, finally obtaining the energy density (now
$D_k = \partial_k - i g A_k$)
\begin{align}
  \label{eq:EtcGL}
   {\density E}  &=
   \tfrac{1}{2} \norm{\vec{D} \psi_\alpha}^2 
    +V\bigl(\psi_{1},\psi_{2}\bigr) + \tfrac{1}{2} \norm{\vec{B}}^2,
\end{align}
The remainder of this paper will be in natural units.

The embedding of Babaev {\it et~al.} \cite{Babaev:2001zy} is such that a closed vortex can be defined by the
fields $\psi_\alpha$, leaving the gauge field $\vec{A}$ free. Using the new variables thus introduced, one can
define a vector field $\vec{n}$ as follows. Let $\vec{\sigma}$ be the usual Pauli matrices. We then define
\begin{align*}
  \vec{n} \defin \begin{pmatrix} \psi_1^* & \psi_2^* \end{pmatrix}
  \vec{\sigma}
  \begin{pmatrix} \psi_1\\ \psi_2 \end{pmatrix}
  = \begin{pmatrix}
    \psi_1^*\psi_2 +\psi_1\psi_2^* \\
    i(\psi_1^*\psi_2 -\psi_1\psi_2^*) \\
    \abs{\psi_1}^2 - \abs{\psi_2}^2
  \end{pmatrix},
\end{align*}
where we demand that $\abs{\Psi}>0$ everywhere, consider $\Psi$ normalised to unity and
$\lim_{\vec{x}\rightarrow \infty} \vec{n}=\vec{n}_\infty$ exists in order to obtain a map $\spac{S}^3
\rightarrow \spac{S}^2$. Now the preimage of $-\vec{n}_\infty$ forms a closed loop, the vortex core.

The fact that $\vec{A}$ is left free, means that there is no nontrivial topology imposed on it and since in
the vacuum of the \GL model $\vec{A}$ is pure gauge, the magnetic field energy can vanish in all cases. This
in turn means that there is no longer a fourth-order derivative in the energy density \eqref{eq:EtcGL} and
Derrick's theorem \cite{Derrick:1964ww} states that no stable, topologically non-trivial solutions of the field
equations with non-zero energy exist. Therefore, the collapse of the magnetic field must be somehow prevented
in order to obtain stable topologically non-trivial configurations in the model. There are several physical
arguments that suggest there might exist physical processes that prevent the collapse, but here we follow the
path set out by Ward, who used a geometrical argument, by adding into the Lagrangian the term (we denote
$\adjoint{\Psi} = (\psi_1^* \quad \psi_2^*)$):
\begin{align}
  \label{eq:Wardterm}
  \density{L}_{W} = \half \kappa^2 \norm{\adjoint{\Psi} D_\mu \Psi}^2
\end{align}
which makes the \GL-Ward energy density
\begin{align}
  \label{eq:EGLW}
  \begin{split}
  \density{E}_{GLW}(x) &=
    \overbrace{\tfrac{1}{2} \norm{\vec{D} \Psi}^2}^{\equiv \density{E}_K} + 
    \overbrace{\tfrac{1}{2} \norm{\nabla \crossprod \vec{A}}^2}^{\equiv \density{E}_B} \\
    &\quad + \overbrace{\half \kappa^2 \norm{\adjoint{\Psi} \vec{D} \Psi}^2}^{\equiv \density{E}_W} +
    \overbrace{V\bigl(\psi_{1},\psi_{2}\bigr)}^{\equiv \density{E}_P}
  \end{split}
  \intertext{and denoting for any subscript $z$: $E_z = \int \diff^3x \density{E}_{z}$ we finally have the
    total energy}
  E_{GLW} &= E_K + E_W + E_B + E_P.
\end{align}
The extra term, when the parameters $\kappa,\eta \rightarrow \infty$, ensures that the model becomes exactly
the Faddeev-Skyrme model
\begin{align}
  \label{eq:EFS}
  \density{E}_{FS} &= \half \norm{\partial_k \vec{n}}^2 + \half g_{FS} \norm{\vec{n} \scalarprod \partial_j
  \vec{n} \crossprod \partial_k \vec{n}}^2
\end{align}
and therefore the model supports, at least asymptotically, stable topologically non-trivial configurations;
these solutions are called knot solitons due to their general shape. This limit of $\kappa \to \infty$ was
apparently first observed by Hindmarsh \cite{Hindmarsh:1992yy}, albeit in a slightly different context.

Ward studied the question whether the solutions of the limiting model remain stable at finite values of
$\kappa,\eta$. It was found, that if $\eta = \kappa^2 + 1$, there are knot solitons already at $\kappa =
7.1$. This is due to the fact that the extra term prevents the (total) collapse of the magnetic field, but
only when $\kappa,\eta$ are large enough: for smaller values, no solutions were found in
Ref.~\cite{Ward:2002vq}, although one was found in Ref.~\cite{Niemi:2000ny}.

It was recently discovered independently by Babaev \cite{Babaev:2008aa} by using physical arguments and by
Speight \cite{Speight:2008na, Speight:2009PC1} by giving a rigorous mathematical proof, that the energy of the
model has no topological lower bound, even when $\density{L}_{W}$ is added, but instead for all values of the
Hopf invariant, the energy can go to zero. Therefore any stable configurations found are necessarily only
local minima of the energy; on the other hand, although the plain \GL model does not seem to have any stable
topologically non-trivial configurations, they may only be very difficult to find due to very small attraction
basin of said configurations, thus requiring very good initial guesses. Investigating such configurations in
closely related models may help in finding these initial configurations.

In Ref.~\cite{Ward:2002vq} Ward investigated only configurations, where $g=1$ and $\eta = \kappa^2 + 1$. We
will now present results for the stable/unstable boundary $\eta(\kappa)$ for $g=1$ and $\kappa \in
\{6,20\}$. It was natural to start the investigation from the values explored by Ward and expand the
range of $\kappa$ in both directions; the limits of this range were eventually set by available computer
capacity.

It is also worth noting, that, as usual, Derrick's theorem provides a virial theorem for the model. Assume we
have a solution of the field equations, $\Psi, \vec{A}$ and consider its energy density under uniform scaling
of the coordinates $x \rightarrow \gamma x$:
\begin{align}
  \label{eq:EDerrick}
  E_{GLW}(\gamma) &= \int \diff^3x \bigl( \density{E}_K(\gamma x) + \density{E}_W(\gamma x) + 
                      \density{E}_B(\gamma x) + \density{E}_P(\gamma x) \bigr)
\end{align}
which is the starting point of Derrick's theorem. Now, following the method used by Derrick, we get, after
the change of integration variables $\vec{x} \rightarrow \gamma \vec{x}$,
\begin{align}
  \label{eq:Derrick_condition}
  E_{GLW}\bigl(\gamma\bigr) &= \gamma E_K + \gamma^{-1} E_B + \gamma E_W + \gamma^3 E_P
  \intertext{differentiating with respect to $\gamma$ we get}
  \gamma^{-2} E_B &= E_K + E_W + 3 \gamma^2 E_P.
\end{align}
The stability under scaling requires that this equation holds for $\gamma=1$, since otherwise some other size
would be energetically more favorable. Thus we have a virial theorem:
\begin{align}
  \label{eq:virial}
  E_B &= E_K + E_W + 3 E_P.
\end{align}
This must hold for all stable minimum energy configurations, regardless of the parameter values or value of
Hopf invariant (topological charge). For the remainder of the article, the energy is rescaled by 
\begin{align*}
  E = \frac{E_{GLW}}{4\pi^2\sqrt{2}}
\end{align*}
and we shall always use $E$ for energy. This rescaling is motivated by Ref.~\cite{Ward:1998pj} and eases
comparisons with Ref.~\cite{Ward:2002vq}.

\section{\label{sec:nummeth}Numerical methods}

We have discretized the system using single-step forward differences on a rectangular cubic lattice. Since the
energy of the \GL model is equal to that of the Abelian Higgs model, the discretization method is
the standard (dropping the time-dependent part) used for lattice quantum field theories, as described in
Refs.~\cite{Wilson:1974sk, Damgaard:1988ec, Jaykka:2006gf}.

The use of single step in the finite differences approximation instead of some more sophisticated alternative
with multiple points is simply a trade-off between speed and accuracy. The discretized equations are very long
even with single step differences. This does not incur significant inaccuracy to the computation for two
reasons. First, we are interested mainly in the existence of knotted solitons, which is not affected by the
less accurate approximation - the exact values of the parameters where the transition from stable to unstable
domain, however, do suffer from inaccuracies as shall be described later. Second, there is no accumulation of
error during the iterative process in the optimization algorithms employed.

The term $\density{E}_W$ was discretized in the same manner as the kinetic term: we denote $\vec{\mu_j^l} =
(\delta_j^1, \delta_j^2, \delta_j^3)\nu^l$, $\nu^l \in \{0,1\}$ and find all gauge invariant discrete terms of
the forms 
\begin{align*}
  &\psi_1^*(\vec{x} + \vec{\mu_j^0}) \psi_1(\vec{x} + \vec{\mu_j^1}) \quad \text{and}\\
  &\psi_2^*(\vec{x} + \vec{\mu_j^2}) \psi_2(\vec{x} + \vec{\mu_j^3}),
  \intertext{where $\nu^0+\nu^2=\nu^1+\nu^3$, $\sum_l\nu^l<4$ and}
  &\psi_1^*(\vec{x} + \vec{\mu_j^0}) \psi_1(\vec{x} + \vec{\mu_j^1}) \quad \text{and}\\
  &\psi_2^*(\vec{x} + \vec{\mu_j^2}) \psi_2(\vec{x} + \vec{\mu_j^3}) e^{\pm i a g A_k(\vec{x})},
\end{align*}
where $\sum_{j=0}^3 \nu^l = 1$.
There are 18 such terms, which are then multiplied by such constants that the sum of the
multiplied terms has the correct continuum limit. This produces the most general single-step
forward-differences discretization of $\density{E}_W$.

Energy minimization was done using several different gradient-based optimization methods: steepest descents
(SD), Fletcher-Reeves (FR) \cite{Fletcher:1964aa} and Polak-Ribi\`ere (PR) \cite{Polak:1969aa} versions of the
conjugate gradient method and Broyden-Fletcher-Goldfarb-Shanno (BFGS) quasi-Newton method
\cite{Broyden:1970aa,Fletcher:1970aa,Goldfarb:1970aa,Shanno:1970aa}. Of these a few simple speed tests were
conducted, with the result that the FR method is usually fastest, but sometimes the SD method is faster due to
the fact that the other methods spend too much time performing the line searches. (The effect of using less
accurate line searches was not investigated.) All the methods are based on gradient directions and thus will
only provide a local minimum. Some of the final configurations were subjected to simulated annealing in order
to see how deep the minimum is. The annealing could not escape the minimum in a reasonable amount of time,
making it reasonable to believe the minima are relatively deep. We did not have the computational resources to
perform simulated annealing optimizations of all cases due to the extreme slowness of the algorithm.

As a test of the accuracy and validity of the programs used, we reproduced the results of
Ref.~\cite{Ward:2002vq} section III. The results agree to within 5\%, where our energies are always higher;
this confirms the correctness of the program and also gives some indication as to the accuracy compared to
other methods. The accuracy of our method could be increased by using larger lattices and smaller lattice
constants, but this kind of brute force approach would require excessive amounts of memory - we use more
than half a terabyte at maximum - so some more sophisticated methods would be needed.

\section{\label{sec:results}Results}

The search for the boundary between stable and unstable domains of $(\kappa,\eta)$ was done as follows. First,
an initial state was set up in a cubic $180^3$ lattice with a lattice constant of $1/18$. The initial
configuration was constructed so that none of the $xyz$-axes coincides with the axial symmetry of the
soliton. This configuration was then minimized using one of the above algorithms and various combinations of
$(\kappa,\eta)$ to determine the rough shape of the boundary. This initial search is done in a small lattice
in full knowledge that it may not be large enough to accurately distinguish between configurations which are
truly unstable and those that are unstabilized due to the large value of the lattice constant. Indeed, all the
unstable systems at the boundary found in this initial search were later proven to be stable in a more
accurate lattice. The search consisted of 72 computer runs, but due to the small lattice, used relatively
little computer time and gave us the rough values around which to start searching for the boundary in a more
accurate lattice. The number of these computationally much more expensive runs was 81.

The values of $(\kappa,\eta)$ were then refined in lattices ranging from $360^3$ to $600^3$. Some unstable
configurations were also put into lattices of sizes up to $720^3$ to support the conclusion that the
instability is real and not caused by discretization effects. None of these were thus stabilised. The virial
theorem Eq.~\eqref{eq:virial} was then checked for the stable configurations at the boundary (i.e. for each
$\kappa$ the stable configuration with lowest $\eta$) to see if the configuration really is a
solution. Allowing for a 10\% inaccuracy, those that were within the tolerance were considered solutions of
the energy minimisation. These points are used to sketch the boundary of the stable configurations. Those that
were outside the tolerance were further investigated.  The reason for inaccuracy proved usually to be due to
the small physical dimensions of the computational lattice: the knot suffers from pressure exerted by the
edge of the box and cannot reach its preferred size. These configurations were therefore put into a larger
lattice with the same lattice constant, minimized and the accuracy was checked again. This process was
repeated as many times as necessary to achieve the desired accuracy - except for two cases as we will describe
later. Whenever the accuracy was reached, the configuration was considered a solution and added to those used
to sketch the boundary. For some cases the accuracy was simply a question of lattice constant; these were
recomputed with same physical dimensions but a smaller lattice constant in order to reach the desired
accuracy.

The exceptional cases where the process of putting into larger box until accuracy is achieved was not
completed, were the pairs $(\kappa,\eta) \in \{(10,0.2), (20,0.14)\}$. These are stable configurations, but the
accuracy goal could not be achieved with the computational capability available due to the cubic growth of
memory requirements of increasing the lattice size.

\begin{figure}[h]
  \centering
  \includegraphics[width=7cm]{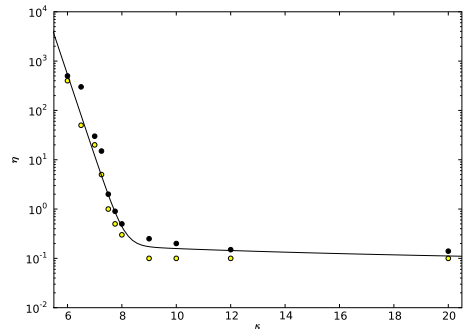}
  \caption{Boundary between stable and unstable regions: the solid black circles denote pairs of $(\kappa,
    \eta_\kappa)$, yellow circles denote the largest unstable values of $\eta$ and the curve, $\eta =
    45^{7.65-\kappa}+0.5 \kappa^{-\half}$, is a sketch of the boundary.}
  \label{fig:boundary}
\end{figure}

After completing the above process, we select for each $\kappa$ the stable configurations with lowest $\eta$;
denote this value by $\eta_\kappa^{\text{min}}$. The values $(\kappa, \eta_\kappa^{\text{min}})$ are displayed
as solid black circles in Fig.~\ref{fig:boundary} together with a curve $\eta(\kappa)$ sketching the
approximate shape of the continuous boundary and yellow circles for largest unstable values of
$\eta$. Configurations for different values of $\kappa$ are not always produced in a lattice of the same size,
but despite that, the boundary curve fits rather well. All the stable dots in Fig.~\ref{fig:boundary} are
confirmed to be solutions by the virial theorem of Eq.~\eqref{eq:virial}, except the cases mentioned above:
$\kappa \in \{10,20\}$. The boundary approaches $y$-axis as $\kappa \rightarrow 0$ and $x$-axis as $\kappa
\rightarrow \infty$. The latter information is not very useful in constructing initial states for normal \GL
model, but the fact that the boundary seems to approach $y$-axis as well, might provide helpful insight and
allow the construction of an initial state which could be used to find a topologically stable, non-trivial
local energy minimum.

\begin{figure}[h]
  \centering
  \includegraphics[width=7cm]{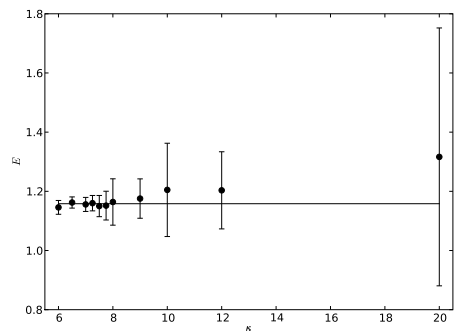}
  \caption{Energies (black discs) of the stable configurations closest to the boundary, the heights of the
    error bars show the inaccuracy of the solution as determined from the virial theorem and the horizontal
    line is the least-squares fit of a constant energy.}
  \label{fig:boundary_energy}
\end{figure}

Comparison of the energies of the final configurations reveals that the energy of the solution for $(\kappa,
\eta_\kappa)$ is, within our numerical accuracy, independent of $\kappa$. This is displayed in
Fig.~\ref{fig:boundary_energy}, where the dots depict the energies of the solutions, the heights of error bars
are chosen according to how much the solution deviates from the virial theorem Eq.~\eqref{eq:virial} and the
solid horizontal line is the least-squares fit for the constant energy, neglecting the the anomalous cases
$\kappa \in \{10,20\}$. Also, comparing this energy with the energies of the unstable configurations at the
moment of loss of topology, shows that the unstable configurations always have lower energy than those at the
boundary, giving even further support to our argument that the instability is real and not a numerical
artefact.

\begin{figure}[h]
  \centering
  \subfloat{\includegraphics{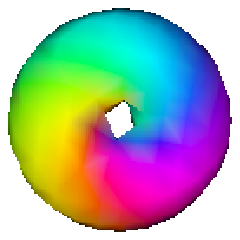}} \quad 
  \subfloat{\includegraphics{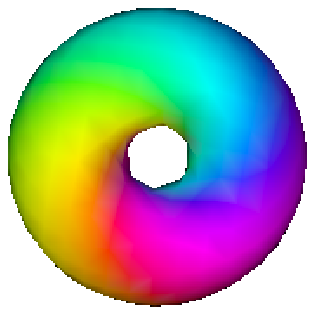}} \\
  \subfloat{\includegraphics{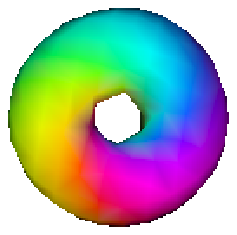}} \quad
  \subfloat{\includegraphics{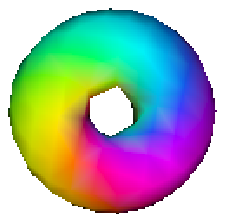}}
  \caption{Initial configuration (top left) and three resulting final configurations: $\kappa=6$ (top right),
    $\kappa=8$ (bottom left) and $\kappa=12$ (bottom right). The coloured region is the equator of $\vec{n}$
    (i.e. isosurface where $n_3=0$) and the coloring corresponds to the longitude of $\vec n$. (Colors
    available on-line.)}
  \label{fig:initial_final}
\end{figure}

The search for the boundary also provides us, as a by-product, with information on the shape and size of the
final configurations. All solutions have kept their initial orientation and overall shape, the only visible
difference between the final configurations is the apparent decrease of the size of the torus with growing
$\kappa$, as shown in Fig.~\ref{fig:initial_final} for the cases $\kappa \in \{6,8,12\}$. It remains an open
question whether the final toroidal configuration obtained from this initial configuration would shrink to
zero as $\kappa \rightarrow \infty$ because we were unable to follow the boundary above $\kappa=12$.

Looking at the four terms of the energy function of these final configurations reveals more details of the
interplay between the various terms. As can be seen from Table~\ref{tab:energies}, $E_K$ increases with
increasing $\kappa$, but $E_W$ decreases. This is expected since on the limit $\kappa \rightarrow \infty$, we
must have $E_W \rightarrow 0$. Since it was found in \cite{Jaykka:2006gf} that the magnetic field always
approaches to zero for unstable systems in non-modified \GL model, it is interesting to note that there seems
to be no trace of this here: the magnetic energy does not change appreciably. The variation in the potential
energy is also negligibly small.

\begin{table}[h]
  \centering
  \begin{tabular}{|c|c|c|c|c|}
    \hline
    $\kappa$ & $E_{K}$ & $E_{B}$ & $E_{W}$ & $E_{P}$ \\
    \hline
    6        & 23.740 & 32.111 & 7.6571 & 0.46053 \\
    \hline
    8        & 26.739 & 31.808 & 5.9745 & 0.46335 \\
    \hline
    12       & 31.580 & 32.334 & 2.5649 & 0.70555 \\
    \hline
  \end{tabular}
  \caption{Terms of the energy function of the final configurations of $\kappa \in \{6,8,12\}$.}
  \label{tab:energies}
\end{table}

\section{\label{sec:conclusions}Conclusions}

We have studied the existence of local minima in a modified two-component \GL model and how the energy of
these behaves along the boundary where the local minima become unstable. It was found that local minima exist
for a wide range of values of the parameter $\kappa$, but that there is a limiting value of $\eta$ for each
$\kappa$ below which the minimum vanishes. It remains open whether there still is a minimum but our initial
configuration has simply moved ``closer'' to the global minimum of zero so that the gradient-based algorithms
can no longer reach it. Also, there can be other local minima. To explore these possibilities further it is
required to use either a set of very different initial configurations or an algorithm which can explore a
wide region of the configuration space starting from a single initial configuration such as the genetic
algorithm (which has the downside of being able to escape the local minima and thus ending up in the trivial
global minimum).

Strikingly, the energy was shown to be constant along the boundary. This information, along with the fact that
higher values of $\eta$ are required for lower values of $\kappa$, might provide a way to construct an initial
configuration in the ordinary two-component \GL model, which, under minimisation of energy, would
lead to a non-zero local minimum. The procedure would, however, require further insight into how the various
terms of the energy functional behave when $\kappa$ decreases;
our numerical scheme was not designed for this and as such, appears not to be accurate enough to provide this
information. In contrast to its energy, the size of the minimum energy configuration decreases as $\kappa$
increases. This requires progressively smaller values of the lattice constant and thus significantly different
numerical approach than the simple, but very large (recall that we used over half a terabyte) lattices used
here. Still, the possibility remains of further research in both smaller and larger values of $\kappa$, but as they
are not addressable by the framework used here, it falls outside the scope of this paper.

\begin{acknowledgments}
  The author wishes to thank Jarmo Hietarinta, Petri Salo, Egor Babaev and Richard Ward for useful discussions
  and Martin Speight for pointing out the non-existence of non-zero topological energy bound.  This work has
  been partially supported by a grant from the Jenny and Antti Wihuri Foundation and partially by a research
  grant from the Academy of Finland (project 123311). The author acknowledges the generous computing resources
  of CSC~--~IT Center for Science Ltd, which provided the supercomputers used in this work.
\end{acknowledgments}

\bibliography{bibliography}

\end{document}